\documentclass{article}

\usepackage{graphicx}
\usepackage{tabularx}
\usepackage{booktabs}

\usepackage[numbers]{natbib}
\newcommand{\LU}{\mathrm{LU}}
\usepackage{siunitx}
\DeclareSIUnit\parsec{pc}
\DeclareSIUnit\lightyear{ly}
\DeclareSIUnit\year{yr}

\usepackage{pbox}

\begin{document}

\title{Dissolving the Fermi Paradox}

\author{
Anders Sandberg, Eric Drexler and Toby Ord\\
Future of Humanity Institute, Oxford University
}


\date{\today}




\maketitle

\begin{abstract}
The Fermi paradox is the conflict between an expectation of a high {\em ex ante} probability of intelligent life elsewhere in the universe and the apparently lifeless universe we in fact observe. The expectation that the universe should be teeming with intelligent life is linked to models like the Drake equation, which suggest that even if the probability of intelligent life developing at a given site is small, the sheer multitude of possible sites should nonetheless yield a large number of potentially observable civilizations. We show that this conflict arises from the use of Drake-like equations, which implicitly assume certainty regarding highly uncertain parameters. We examine these parameters, incorporating models of chemical and genetic transitions on paths to the origin of life, and show that extant scientific knowledge corresponds to uncertainties that span multiple orders of magnitude. This makes a stark difference. When the model is recast to represent realistic distributions of uncertainty, we find a substantial {\em ex ante} probability of there being no other intelligent life in our observable universe, and thus that there should be little surprise when we fail to detect any signs of it. This result dissolves the Fermi paradox, and in doing so removes any need to invoke speculative mechanisms by which civilizations would inevitably fail to have observable effects upon the universe.
\end{abstract}




\section{Introduction}
While working at the Los Alamos National Laboratory in 1950, Enrico Fermi famously asked his colleagues: "Where are they?" \cite{joneseverybody}. He was pointing to a discrepancy that he found puzzling: Given that there are so many stars in our galaxy, even a modest probability of extraterrestrial intelligence (ETI) arising around any given star would imply the emergence of many such civilizations within our galaxy. Further, given modest assumptions about their ability to travel, to modify their environs, or to communicate, we should see evidence of their existence, and yet we do not. This discrepancy has become known as the {\em Fermi paradox}, and we shall call the apparent lifelessness of the universe the {\em Fermi observation}.

Many hypotheses have been suggested in efforts to resolve the Fermi paradox, for example, that all other civilizations are deliberately concealing themselves, or that they all annihilate themselves before successfully traveling or communicating at interstellar distances. A major difficulty for such hypotheses is that the putative mechanism must be extremely reliable: If only \SI{99}{\percent} of other civilizations annihilated themselves, this would do little to resolve the paradox. These hypotheses are thus highly speculative, relying on strong implicit claims about universal alien motivations or social dynamics, when we cannot claim similar knowledge of our own world. These hypotheses are entertained not out of independent scientific plausibility, but because they are seen as potential explanations of he Fermi observation.

Our main result is to show that proper treatment of scientific uncertainties dissolves the Fermi paradox by showing that it is not at all unlikely {\em ex ante} for us to be alone in the Milky Way, or in the observable universe. Our second result is to show that, taking account of observational bounds on the prevalence of other civilizations, our updated probabilities suggest that there is a substantial probability that we are alone. Our third result is that pessimism for the survival of humanity based on the Fermi paradox is unfounded.


\subsection{The Drake Equation}
The key assumption of the Fermi paradox is that the number of sites where alien civilizations could emerge is so large that for any reasonable probability of emergence, some would have emerged and we should expect to have detected one or more of them. This {\em sites} $\times$ {\em probability} approach fits within the well-known Drake equation framework, which we shall take to be the paradigm example of this form of reasoning about the prior probability of ETI.

The Drake equation was intended as a rough way to estimate of the number of detectable/contactable civilizations in the Milky Way ($N$) \citep{drake2015drake}, phrased as a product of seven factors\footnote{While there exist various similar formulations, we will focus on the classic Drake Equation. Those who prefer the other formulations should have no great difficulty transferring our conclusions to their preferred framework.}:

$$N=R_*f_pn_ef_lf_if_cL$$ 

Where:
$R_*$ is the rate of star formation per year,
$f_p$ is the fraction of stars with planets, 
$n_e$ is the number of Earth-like (or otherwise habitable) planets per system with planets,
$f_l$ is the fraction of such planets with life, 
$f_i$ is the fraction with life that develop intelligence, 
$f_c$ is the fraction of intelligent civilizations that are detectable/contactable, and 
$L$ is the average longevity of such detectable civilizations in years. 

The Drake equation has been a mainstay in the SETI debate, sometimes being used to directly estimate the number of civilizations in the galaxy, but perhaps more often being used as an analysis tool. For example, it has played a prominent role in debating the rationality of SETI efforts. This approach to the Drake equation is well summed up by Jill Tarter, who said "The Drake Equation is a wonderful way to organize our ignorance" \cite{Achenbach2000}.

But while the equation is often invoked as a way of reasoning about uncertainties and ignorance, the actual practice is often considered to be somewhat suspect. Many papers state that some of their parameter choices are just their best guesses, though this fails to provide an appropriate framework for interpreting the result. It is common to see carefully estimated astrophysical numbers multiplied by these ad hoc guesses. It has been noted that the final results seem to depend heavily on the pessimism or optimism of the authors, falling into the "$N \approx L$" and "$N \approx 1$" schools respectively. Steven J. Dick provides a typical statement of this worry: "Perhaps never in the history of science has an equation been devised yielding values differing by eight orders of magnitude. ... each scientist seems to bring his own prejudices and assumptions to the problem." \cite{Dick}

Nevertheless, an equation showing the degree of divergence in estimates can be very valuable. The problem is when systematic biases and assumptions dominate the results. As we will see, the fact that answers {\em only} span eight orders of magnitude appears to be due to overconfidence -- the range should be substantially wider.

While all practitioners acknowledge the great uncertainty around the parameters of the Drake equation, very few incorporate this into their quantitative models. The Drake equation (and related models) are almost always used with point estimates for each parameter, rather than ranges or probability distributions. If the result is used to estimate the chance of ETI in our galaxy (as is common when introducing the Fermi paradox), this can be extremely misleading. It is not enough to claim that the output of the equation is just an order of magnitude estimate: scientific uncertainty about the parameters does not constrain the output to one or even a handful of orders of magnitude. In our view the practice of using point estimates in Drake equation-like frameworks is largely responsible for the continued puzzlement about the Fermi paradox.

\subsection{Knowledge claims of point estimates}

It is instructive to ask what knowledge claims about the parameters are implicit in the use of point estimates. The answer is that they implicitly claim complete certainty about all the parameters. We shall see that it is this presupposition of certainty that is creating the appearance of a paradox (by falsely representing how certain we are that there are many civilizations out there). In this paper we provide a principled approach to assessing the reasonable uncertainty in each parameter via consideration of what knowledge is being claimed, and to tracking this uncertainty through the calculation, showing how this proper management of our uncertainties dissolves the apparent paradox.

To quickly see the problems point estimates can cause, consider the following toy example. There are nine parameters $(f_1, f_2, \ldots)$ multiplied together to give the probability of ETI arising at each star. Suppose that our true state of knowledge is that each parameter could lie anywhere in the interval $[0, 0.2]$, with our uncertainty being uniform across this interval, and being uncorrelated between parameters. In this example, the point estimate for each parameter is 0.1, so the product of point estimates is a probability of 1 in a billion. Given a galaxy of 100 billion stars, the expected number of life-bearing stars would be 100, and the probability of all 100 billion events failing to produce intelligent civilizations can be shown to be vanishingly small: \num{3.7e-44}. Thus in this toy model, the point estimate approach would produce a Fermi paradox: a conflict between the prior extremely low probability of a galaxy devoid of ETI and our failure to detect any signs of it. 

However, the result is extremely different if, rather than using point estimates, we take account of our uncertainty in the parameters by treating each parameter as if it were uniformly drawn from the interval $[0, 0.2]$. Monte Carlo simulation shows that this actually produces an empty galaxy \SI{21.45}{\percent} of the time: a result that is easily reconcilable with our observations and thus generating no paradox for us to explain. That is to say, given our uncertainty about the values of the parameters, we should not actually be all that surprised to see an empty galaxy. The probability is much higher than under the point estimate approach because it is not that unlikely to get a low product of these factors (such as 1 in 200 billion) after which a galaxy without ETI becomes quite likely. In this toy case, the point estimate approach was getting the answer wrong by more than 42 orders of magnitude and was responsible for the appearance of a paradox. 

In this paper, we shall look at two different ways of extending this approach beyond a toy model --- generating probability distributions for the parameters of the Drake equation based on the variation in historical estimates and doing so based on the authors' best judgment of the scientific uncertainties for each parameter. In both cases, we see an effect like the one in this toy model, making a galaxy (or observable universe) without ETI quite plausible {\em ex ante} and thus dissolving any apparent paradox.

It should be noted that there do exist some cases in the literature using estimated uncertainties rather than point estimates, such as treating each factor as uniformly distributed in an uncertainty interval and then convolving them into a final distribution \cite{maccone2011seti}, or treating the system as a stochastic process \cite{glade2012stochastic} or Monte Carlo simulation \cite{forgan2009numerical}. The probability of life emerging has also been studied in a Bayesian framework \cite{spiegel2012bayesian}. While these papers aim at improving the precision of the Drake equation they do not apply their conclusions directly at resolving the Fermi paradox. It is also possible to use the Drake equation to derive bounds on the parameters using the Fermi observation \cite{prantzos2013joint}.

\section{Using variation in point estimates to model uncertainty}

The literature on the Drake equation contains dozens of point-estimate based calculations for $N$ (the number of detectable civilizations in our galaxy).\footnote{The estimates can be found in Supplement III.} These estimates span 11 orders of magnitude: from \num{3e-4} to \num{1e8}. While each estimate does not contain information about its uncertainty, one could use the variation in the estimates as a proxy for uncertainty in the result. We can do even better if we decompose these estimates into the estimates for each of the parameters and then recombine them in different ways. This gives us a feeling for how likely we should find it that new estimates would fall outside the range of existing estimates, and how far outside this range they might fall.

\begin{figure}
\centering
\includegraphics[width=1.0\linewidth]{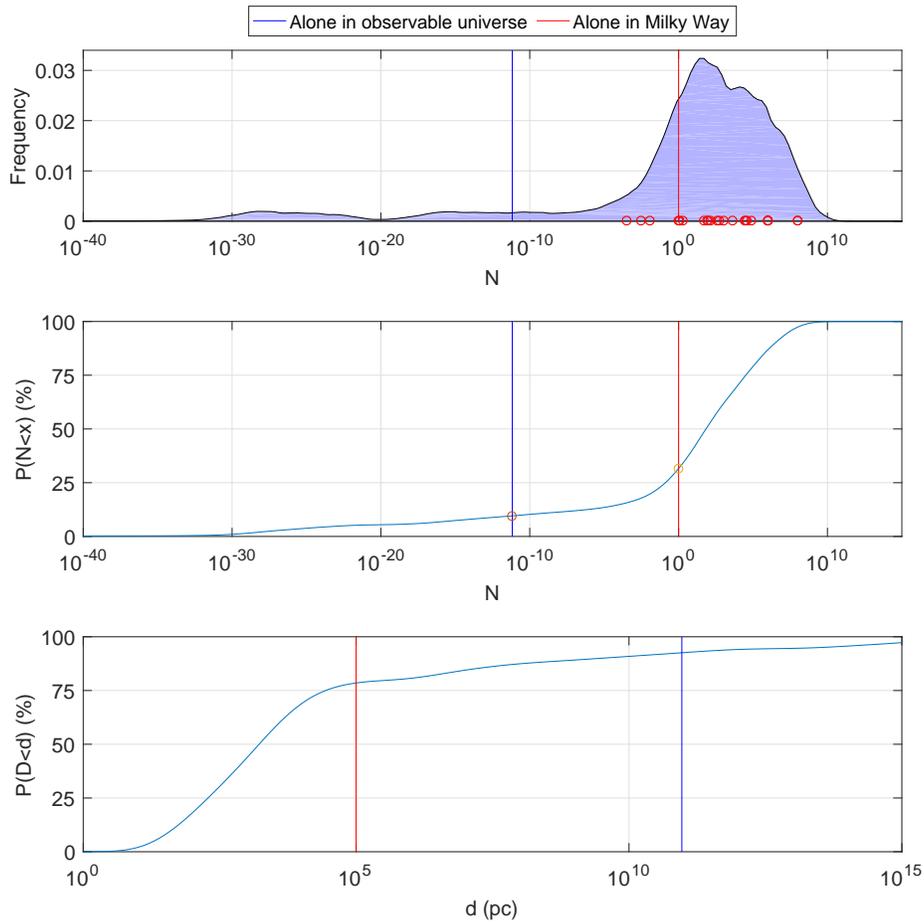}
\caption{(A) The uncertainty in the research community represented via a synthetic probability density function over $N$ --- the expected number of detectable civilizations in our galaxy. The curve is generated by random sampling from the literature estimates of each parameter. Direct literature estimates of the number of detectable civilizations are marked with red rings. (B) The corresponding synthetic cumulative density function. (C) A cumulative density function for the distance to the nearest detectable civilization, estimated via a mixture model of the nearest neighbor functions $F(D<r|N)=1-e^{-4\pi r^3 \rho(r) N/3 N_{MW}}$ where $\rho(r)$ is a sigmoid fit between the star density in the Milky Way and the lower density at larger scales and $N_{MW}=\num{300e9}$ is the number of stars in the Milky Way.}
\label{fig:fig1}
\end{figure}

We do this by generating synthetic point estimates for $N$ by randomly sampling each parameter from the set of estimates for that parameter, and then multiplying these together. Doing this many times produces the following picture of the 'collective' view of the research community's uncertainty about $N$, shown in Fig.~\ref{fig:fig1}.

Of the raw literature estimates for $N$, \SI{64}{\percent} have $N > 100$. As we saw in the toy model, such values of $N$ correspond to probabilities of less than \num{3.7e-44} that we are alone in our galaxy. So if we took any of those estimates at face value, we would face a serious discrepancy with our observations and thus a Fermi paradox. 

However, adjusting for the implicit uncertainty in the literature changes this story dramatically. While most of the synthetic estimates of $N$ are above 1 and the peak of the distribution is on the optimistic side (mean 53 million, median 100), there exists a pessimistic tail due to the existence of some very low parameter estimates. This tail is extremely broad, spanning more than 30 orders of magnitude. 

Given this synthetic distribution for the community's uncertainty, the total credence for $N<1$ is \SI{30}{\percent}. To show that this isn't purely driven by outliers, the bootstrap confidence interval for $N < 1$ is [27\%, 52\%]. This suggests that people who take the views of most members of the research community seriously should ascribe something like a one in three chance to being alone\footnote{It should be noted that an estimate of $N=1$ means that the expected number of civilizations in the Milky Way is 1, not that the actual number is 1. The probability of an empty galaxy given $N$ is $(1-N/N_{MW})^{N_{MW}}\approx e^{-N}$, where $N_{MW}$ is the number of stars in the galaxy. However, for our purposes the difference between $\int_0^\infty e^{-N}P(N)dN$ and $\Pr[N<1]$ is so small (1\% in this case) that we will simply use estimates of $\Pr[N<1]$ as the probability of our being alone in the galaxy. } in the galaxy and so should not be greatly surprised by our lack of evidence of other civilizations. The probability of $N < 10^{-10}$ (such that we are alone in the observable universe) is 10\% (CI [0.2\%, 20\%]). Thus the implicit view of the literature is that being completely alone is distinctly possible, albeit unlikely.

Given a distribution of $N$, we can calculate the distribution of the expected distance to the nearest civilization. Again the synthetic distribution has an optimistic outlook, with a 50\% chance of it being within a kiloparsec, but also a non-trivial probability to the nearest civilization being far beyond the observable universe.

\section{Estimates of current scientific uncertainties}
While the above analysis indirectly estimated uncertainty in the Drake parameters via their variability across the SETI literature, we can also directly estimate some of the uncertainties using domain specific information for each parameter. In particular, in the next section we will show that there are good reasons to assume a far greater uncertainty for $f_l$ and $f_i$ than is common in the SETI literature.

In the following we will commonly be dealing with uncertainties covering many orders of magnitude. Hence it will be convenient to have a way of summarising such degrees of uncertainty. We shall thus define the "log-uncertainty" of a parameter $X$ ($\LU[X]$) as an estimate of the number of orders of magnitude covered by our current, rationally founded, uncertainty. Thus if we say $\LU[X]=3$, we mean that it would be bold to claim that the scientific uncertainty covers less than three orders of magnitude based on our current information.

For an overview of historical estimates of the parameters, see \cite{drake2015drake}.

$R_*$ is fairly well constrained by astronomical data. While star formation rates in other galaxies may vary over 5 orders of magnitude and there has likely been some significant time variation in the Milky Way, the actual current uncertainty is from 2 to 16 solar masses, $\LU[R_*]\approx 0.9$.

Different methods are converging on $f_p \approx 1$. $\LU[f_p]<1$. Estimates of $n_e$ remain far more uncertain, ranging from $<10^{-12}$ in rare earth arguments to $>1$ when taking non-terrestrial environments like icy moons into account. Hence it can be argued that $\LU[n_e]>12$, although the post-2000 literature estimates only covers 6 orders of magnitude. Much of the disagreement is about what requirements go into the $n_e$ term and which ones go into the $f_l$ term. For our purposes we will take an earth-like planet to be be little more than a rocky planet in a habitable zone and thus assume that $\LU[n_e]\approx 2$.

$f_l$ and $f_i$ are highly uncertain, and will be examined in the next section. Similarly there are no clear arguments for the range of uncertainty of $f_c$: the range of estimates in the literature from $10^{-2}$ to 1 give $\LU[f_c]=2$, but it is clear a broader range is intended.

The final $L$ factor ranges between $50<L<10^9-10^{10}$ years, giving $\LU[L]=8.3$. The upper limit occurs because the Drake equation assumes a steady state: even if civilizations survived $10^{11}$ years instead of $10^{10}$ years, this would not increase the number of them found in our galaxy since insufficient time has elapsed for this to make a difference. Thus for practical use of the Drake equation we can cut off the distribution at $10^{10}$ years.

\subsection{Scientific uncertainty in \boldmath$f_l$ and  \boldmath$f_i$}

We shall now provide a sketch of the deep scientific uncertainties regarding $f_l$ and  $f_i$. A substantially more detailed account can be found in supplement I.

As noted by Carter and McCrea \cite{carter1983anthropic} the evidential power of the early emergence of life on Earth is weakened by observer selection effects, allowing for deep uncertainty about what the natural timescale of life formation is. By their argument one cannot assume it lies within the habitability span of Earth.

Following \cite{spiegel2012bayesian}, we model abiogenesis events as physical transitions that occur at some rate per unit time per unit volume of a suitable prebiotic substrate. The probability of observing such an event on a potentially habitable planet with volume $V$ of substrate, during the available period $t$, with an abiogenesis rate $\lambda$ transitions per unit volume and time is $f_l=1-e^{-\lambda Vt}$. For small values, $f_l \approx \lambda Vt$.

We take $t$ to range from  \SIrange{\approx e7}{e10}{\year} depending on whether abiogenesis can happen only during a rare period of global geological change or at any time during an interval of planetary habitability. Thus $\LU[t] \approx 3$.

There is more uncertainty concerning $V$, the volume of prebiotic substrate in which abiogenesis could occur. Abiotic polymerization may be limited to occurring in a thin film of substrate in a region with adequate geothermal heating \cite{lambert2008adsorption} \cite{martin1998free} and productive outcomes may require local concentrations of specific monomers orders of magnitude higher than the average prebiotic levels \cite{luisi2015chemistry}. Each of these issues could reduce $V$ by 10 or more orders of magnitude relative to deeper substrates over a substantial fraction of a planetary surface. We will therefore take $\LU[V] \geq 20$.

There is great uncertainty concerning $\lambda$, the volumetric rate parameter for abiogenesis events. Supplement I takes the frequency of bacterial cell division per bacterial volume as an {\em extremely} conservative upper bound on $\lambda$, and considers potential lower bounds motivated by the so-called Levinthal protein-folding paradox \cite{levinthal1968there} \cite{zwanzig1992levinthal}, which arises in a model of folding that postulates a (counterfactually) random search among alternative peptide backbone conformations. 

In the Levinthal model, rates decline exponentially with protein size, with waiting times $\gg 10^{200}$ times the present age of the universe for the folding of a moderate-size protein. In reality, rapid protein folding is enabled by a "funnel-like" energy surface that is a product of evolution, but a similar result in abiogenic processes would require a different and speculative mechanism of pre-evolutionary self-organization. Exponential scaling in size parameters transforms broad uncertainties in the scale of abiotic to biotic transition states into log-broad uncertainties in transition rates. Here we will fold parameters together and take as a reference value $\LU[\lambda Vt] \geq 200$, though one could easily argue for much larger $\LU$.

Abiogenesis is the first of a chain of major transitions in the development of life comparable to our own. Depending on how one defines ones terms, these could be thought of as contributing to the $f_l$ or $f_i$ term. There is substantial evidence that a so-called "RNA world", in which both genetic and metabolic roles were filled by RNA, preceded the current genetic system of DNA, RNA, and ribosomally translated proteins \cite{bernhardt2012rna}. The genetic transition from an RNA-world biology (or a similar system) to a translation-based biology comparable to our own may be a requirement for the emergence of complex life and intelligence within the time available.

As discussed in Supplement I, several models and lines of evidence suggest that it would be unsurprising to find that alternative genetic systems that either favor or disfavor functional equivalents of our own biology are either very common or very rare. An obvious alternative is between a translation-based biology and a persistent analog of an RNA world; others include enriched-transcript (rather than translation-based) genetic systems, alternative genetic molecules (illustrated by known alternative nucleobases and backbones \cite{engelhart2010primitive,karri2013base,pinheiro2012xna}), and alternatives to uniform, contiguous 3-base genetic codes (again, several have been described \cite{crick1966genetic,hohsaka2002incorporation,baranov2009codon}). Each of these alternatives might (or might not) substantially impede the rate or scope of evolution and hence fail to produce complex life within the time available in the present universe. In each instance, several lines of argument indicate that our uncertainties regarding the associated branching ratios must be regarded as log-broad. More frequently referenced uncertainties downstream from the development of prokaryote-equivalents (e.g., the emergence of eukaryotes, multicellular life, and intelligence itself) contribute further uncertainties.

\subsubsection{The overall uncertainty range for \boldmath$f_l$ and \boldmath$f_i$}

Since our arguments are only strengthened by high uncertainties, we shall conservatively take $\LU[f_l]$ to be 200 and use the uncertainty in $f_i$ based on the literature estimates of 0.001 to 1 (this corresponds to regarding $f_l$ as the fraction of planets with evolutionary competent life rather than just life; see Supplement II for results with equal uncertainty of $f_l$ and $f_i$). We represent the current scientific uncertainty over the rate of abiogenesis per lifetime of a habitable planet with a log-normal, whose standard deviation is 50 orders of magnitude. It is highly unclear where to center this distribution, so we have chosen a location that only makes our argument more difficult: setting the median to the high rate of 1 abiogenesis event per planet. From this $f_l$ is calculated as $f_l=1-e^{-\lambda Vt}$ (giving $f_l$ mean 0.5 and median 0.63).

\subsection{Using these scientific uncertainties in the Drake Equation}
Given the current best estimates and the above argument regarding $f_l$, we can demonstrate the effect of these uncertainties by combining simple uncertainty distributions roughly corresponding to the current state of knowledge.\footnote{In the main text, we treat the Drake equation factors as uncorrelated. While they were intended to represent independent groups of factors, H\"aggstr\"om and Verendel have shown that if there are enough correlations the Great Filter argument is affected. This could occur due to spatial or temporal correlations due to panspermias or gamma-ray bursts. We find that the effect is modest for these two cases (Supplement II).}

%
%
%
%
%
%
%

In the following we will use the simple sketch of the state of current knowledge as in Table \ref{tabparam}\footnote{Log-uniform and log-normal mean that the log of the variable is uniformly/normally distributed.}.

\begin{table}
\centering
\caption{Parameters of simple sketch of current knowledge.}
\small
\begin{tabular}{rl}
\midrule
Parameter & Distribution \\
\midrule
$R_*$ & log-uniform from 1 to 100. \\
$f_p$ & log-uniform from 0.1 to 1.\\
$n_e$ & log-uniform from 0.1 to 1.\\
$f_l$ & log-normal rate, described in previous section.\\
$f_i$ & log-uniform from 0.001 to 1.\\
$f_c$ & log-uniform from 0.01 to 1.\\
$L$ &log-uniform from 100 to 10,000,000,000.\\
\midrule
\end{tabular}
\label{tabparam}
\end{table}

\begin{figure}
\centering
\includegraphics[width=1.0\linewidth]{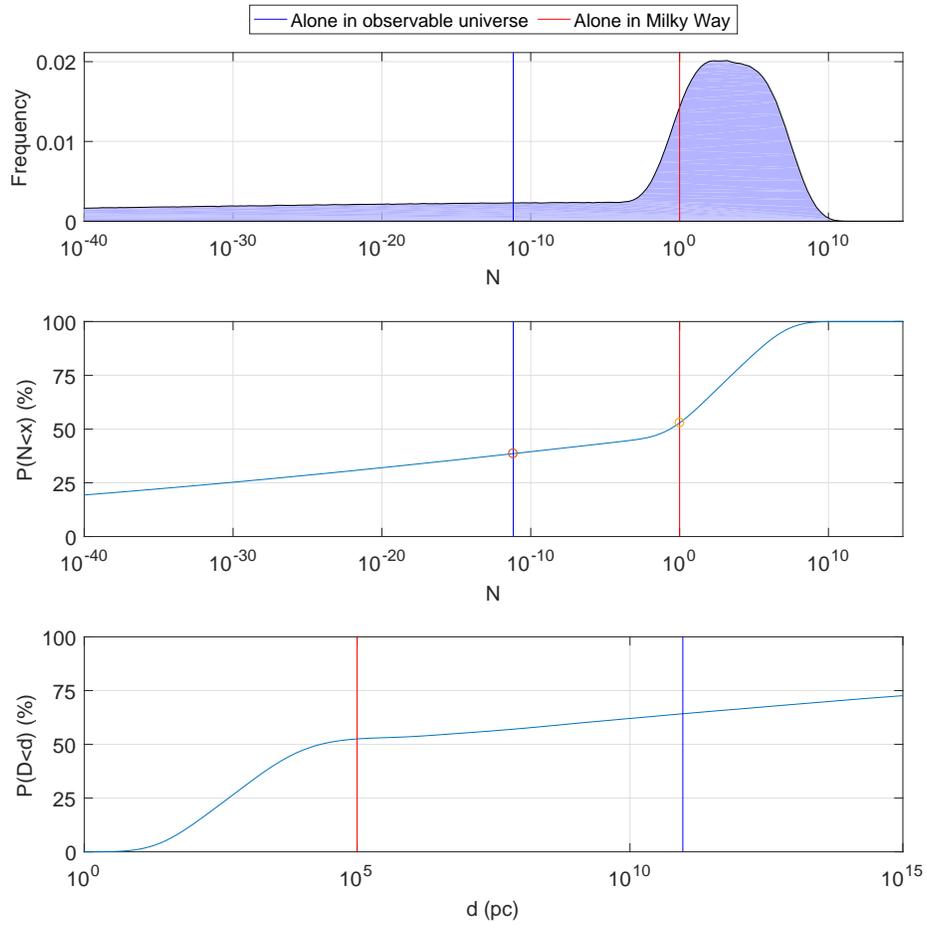}
\caption{(A) A probability density function for $N$ generated by Monte Carlo simulation based on the authors' best estimates of our current uncertainty for each parameter. (B) The corresponding cumulative density function. (C) A cumulative density function for the distance to the nearest detectable civilization.}
\label{fig:fig2}
\end{figure}

The results are shown in Fig.~\ref{fig:fig2}. They are similar to those generated by the literature resampling method, but even stronger. The mean for $N$ is very optimistic, at 27 million, but the median is now only 0.32 -- less than one civilization per galaxy like our own. The probability of $N < 1$ is now \SI{52}{\percent} (up from \SI{30}{\percent} with literature resampling). Most markedly, the very uncertain life formation rate produces a heavy left tail, giving a nearly \SI{38}{\percent} credence that $N < 10^{-10}$, making us alone in the observable universe (up from \SI{10}{\percent} with literature resampling).\footnote{The distribution shape in log-space is close to a mixture of two Gaussians. The multiplicative central limit theorem states that products of finite-variance independent factors tends towards a lognormal distribution as the number of factors increase. However, since the number of factors is just seven the strongly bimodal $f_l$ distribution causes the head-tail mixture shape.}

While the analysis above required us to make our own judgment calls about how to represent the state of scientific uncertainty for each of these parameters, our qualitative result is robust to many of these assumptions and can be driven by our claimed uncertainty in $f_l$ alone. Even if all other parameters are set to the most optimistic point estimates from the literature (with no uncertainty), we still get \SI{41}{\percent} credence that $N < 1$ and \SI{32}{\percent} credence that $N < 10^{-10}$ (see figure S1 in Supplement II). Similarly, even if the estimate for $f_l$ were made much more optimistic, such as by shifting the median value for $\lambda_l$ up by 10 orders of magnitude (so that the median rate of abiogensesis on a habitable planet is in the order of once per year), the effect is minor, since the extremely broad left-hand tail extends far more than 10 orders of magnitude.

We can thus state our first main conclusion. While using point-estimates in the Drake equation frequently generates estimates of $N$ that would produce a Fermi paradox, this is just an artefact of the overconfidence implicit in treating them as having no uncertainty. When our uncertainty is properly accounted for in the model, we find a substantial prior probability that there is no other intelligent life in our observable universe, and thus that there should be little surprise when this is what we see. We find this when using the variation in parameter estimates in the literature as a proxy for our state of uncertainty and when generating our own summaries of the current state of uncertainty in the scientific literature.

Note that this conclusion does not mean that we {\em are} alone (in our galaxy or observable universe), just that this is very scientifically plausible and should not surprise us. It is a statement about our state of knowledge, rather than a new measurement.

\section{Updating on the Fermi Observation}

We have used the scientific uncertainty about the parameters of the Drake equation to construct an {\em ex ante} credence distribution for $N$, but have not yet factored in the 'Fermi observation' --- the lack of any direct evidence of ETI. Updating our credences based on the Fermi observation means lowering our credence in those parameter combinations that produce high likelihoods of abundant ETI. 

There is considerable uncertainty about the range from which a civilization's radiation could be detected. It has been estimated that typical radiation from contemporary humanity could be detected from \SI{18}{\parsec}, while special transmissions such as the Arecibo message could be detected from 30 to \SI{1470}{\parsec}\cite{sullivan1978eavesdropping,billingham1992detection,chela2012astrobiology,tarter2001search}. However, sky surveys may be much less sensitive\cite{tarter2001search}. Scintillation may limit the ability to re-detect radio signals beyond a few hundred parsecs\cite{cordes1997scintillation}. Optical searches could detect megajoule pulses from 10 meter apertures out to a distance of \SI{60}{\parsec}\cite{tarter2001search}. The \^G-survey established that there were no type III Kardashev civilizations using more than \SI{85}{\percent} of the starlight in $10^5$ surveyed galaxies\cite{griffith2015g}.\footnote{With only a small subset of galaxies consistent with \SI{>50}{\percent}.}

Radiation is not the only way we could detect the existence of a civilization. Another possibility is via interstellar travel --- either to our solar system or to a nearby system from which their radiation could be more easily detected. Galactic settlement timescales have been estimated to be below \SIrange{e6}{e10}{\year}, with most estimates on the order of \SI{e7}{\year} \cite{jones1976colonization,newman1981galactic,jones1981discrete,fogg1987temporal,wright2014g}. This leads to the argument that even for very conservative travel speeds, the entire galaxy would be occupied in a very short time and hence the lack of observable settlement provides a strong bound on extraterrestrial intelligence.\footnote{While it is rarely considered in the literature, if intergalactic travel is possible, this could also greatly increase the effective detection range and the bound.}

In our framework, the Fermi observation leads to an update of our credence distribution of $N$ as per Bayes' rule: $$\Pr[N|\neg D]=\frac{\Pr[\neg D|N]\Pr[N]}{\Pr[\neg D]},$$ where $\neg D$ denotes no detection. Which value we should use for $\Pr[\neg D|N]$ depends on our model of the Fermi observation.

\subsubsection{Random sampling update}

If the observation is that out of $K$ sampled stars none have have a civilization, then: $$\Pr[\neg D|N,K]=\left(1-(N/N_{MW})\right)^K$$

\subsubsection{Spatial Poisson update}

If the observation is that there is no civilization closer than some detection distance $d$, then: $$\Pr[\neg D|N,d]=1-e^{-4\pi(N/V_{MW})d^3/3}$$

\subsubsection{Settlement update}

Models incorporating interstellar settlement produce very strong updates, but are also problematic to incorporate in the steady-state Drake equation framework (see Supplement II for more details and several alternative models). 
If the observation is that no nearby spacetime volume has ever been permanently settled, then given a settlement timescale $T$ and an effective age $T_{MW}$ of the Milky Way  the update becomes $$\Pr[\neg D|N] \approx e^{-(N/L)(T_{MW}-T)} + \left(1-e^{-(N/L)(T_{MW}-T)}\right)\frac{\alpha}{\alpha+1},$$ where $2<\alpha<3$ is a geometric factor due to the shape of the galaxy.


\subsubsection{K3 update}

If the observation is that out of $K$ sampled galaxies, none of them have type III Kardashev civilizations, then given a probability $P_{K3}$ that type III civilizations are technologically possible, and a probability $P_{K3}$  that an intelligent civilization succeeds in becoming one:
$$\Pr[\neg D|N,K,P_{K3},P_{succ}]=1-P_{K3}\left(1-(1-P_{succ})^K\right)$$

\begin{figure}
\centering
\includegraphics[width=1.0\linewidth]{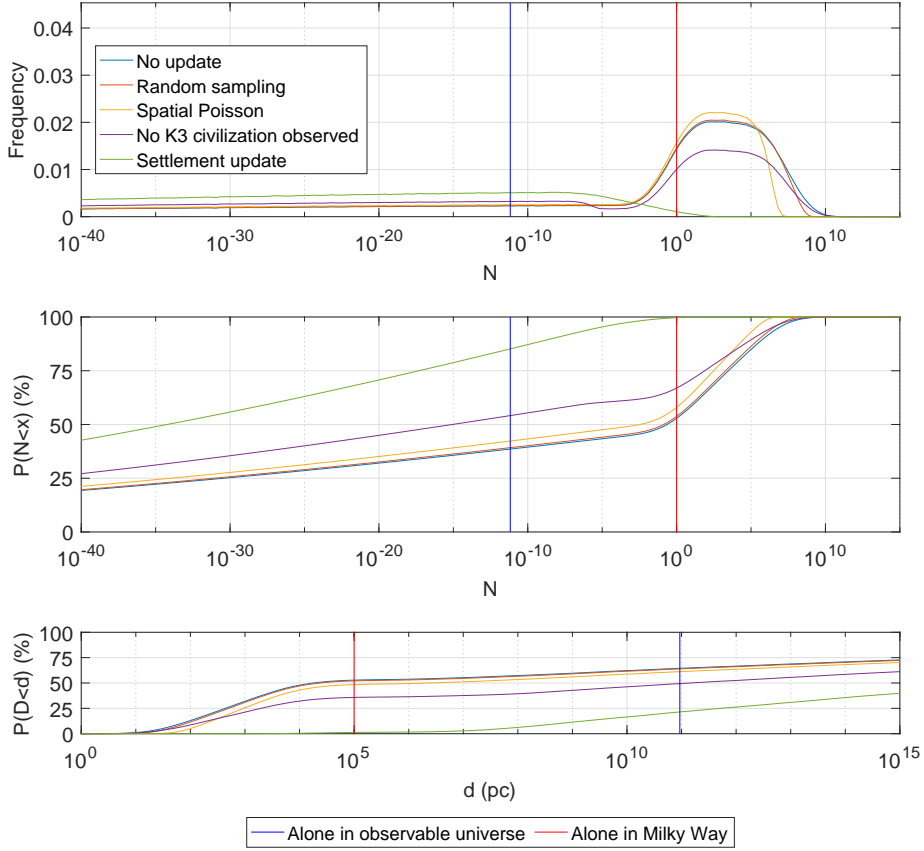}
\caption{Effect of updating on the Fermi observation for different scenarios. Random sampling: no observations among 1,000 studied stars, spatial Poisson: no ETI within 18 pc, no K3 civilization: effect of the \^G-survey if $P_{K3}=0.5$ and $P_{succ}=0.01$, no settlement: no interstellar civilization has yet spread to our location in the galaxy.
}
\label{fig:figpost}
\end{figure}

The effects of conditioning can be seen in Table~\ref{tab:conditioning} and Figure~\ref{fig:figpost}.

\begin{table*}
\centering
\caption{Comparison of conditioned credence distributions.}
\small
\begin{tabular}{p{2cm}rrrrrrrr}
\midrule
Update & Mean $N$ & Median $N$ & $\Pr[N<1]$ & \pbox{30cm}{$\Pr$ \\$[N<10^{-10}]$} &  \pbox{20cm}{Median \\$f_l$} & \pbox{20cm}{Median\\ $L$}\\
\midrule
No update&\num{2.7e+07}&\num{0.32}&\num{0.52}&\num{0.38}&\num{0.64}&\num{1e+06}\\
Random sampling&\num{2.5e+06}&\num{0.19}&\num{0.53}&\num{0.39}&\num{0.09}&\num{8.6e+05}\\
Spatial Poisson&\num{7.8e+04}&\num{0.0048}&\num{0.57}&\num{0.42}&\num{3.1e-06}&\num{4.5e+05}\\
K3 update&\num{1.9e+07}&\num{1.2e-15}&\num{0.66}&\num{0.54}&\num{4e-19}&\num{9e+05}\\
Settlement update&\num{0.072}&\num{8.1e-35}&\num{0.996}&\num{0.85}&\num{3e-38}&\num{1e+06}\\
\midrule
\end{tabular}
\label{tab:conditioning}
\end{table*}

The Fermi observation thus raises our 52\% credence for being alone in the galaxy to somewhere between 53\% and 99.6\%, depending on the type of evidence considered. It likewise raises our 38\% credence for being alone in the observable universe to somewhere between 39\% and 85\%. In both cases the evidence from scanning nearby stars for radiation signals produces changes on the scale of a few percentage points, while the evidence of a lack of interstellar settlement in the Milky Way or other galaxies produces much larger changes.

\subsection{Fallible Fermi observations}

An important caveat is that an apparently negative observation can be due to either actual absence or failure to recognize ETI. This can be due to an individual failure for this particular observation, or a systematic failure at recognizing true ETI.  Individual failures include problems detecting the relevant signal as well as independent choices by the ETIs not to communicate or be visible in that way. Systematic failure can include systematic problems detecting the relevant signal or systematic reasons that the ETIs do not communicate or are not visible in that way (such as it being technically impossible).

Given $K$ independent negative observations with individual failure probability $p_{indiv}$ the update becomes $\Pr[\neg D|N]=(p_{indiv} + (1-p_{indiv})\Pr[\neg d|N])^K$, where $\Pr[\neg d|N]$ is the probability of not detecting ETI in a single accurate observation. Even if $p_{indiv}$ is macroscopic, it can easily be overwhelmed by a large number of observations. 

Given a probability $p_{syst}$ of systematic failure, the update becomes $p_{syst} + (1-p_{syst})\Pr[\neg D|N]$. Thus the true posterior is just a linear interpolation between the prior and posterior one gets when ignoring systematic failure. Given the magnitude of the previous results, unless one assigns a very strong prior to $p_{syst}$ being large, our qualitative conclusions still hold.

\subsection{Related arguments}

Our argument so far is related to a recent argument sketched by Max Tegmark \cite{tegmark2014mathematical}. Like us, he suggests that we should have great uncertainty about $f_l$ and $f_i$, making us very uncertain about the probability of intelligent life arising around a given star. He thus models our uncertainty over the average distance between two independently arising intelligent civilizations as log-uniform. That is, we should be no more surprised if this average distance were at one order of magnitude rather than another. Thus, when we gain some evidence that there is no other civilization within our galaxy, we update this prior by greatly lowering our credence in the average distance being less than this (\SI{\approx e21}{\metre}). Since there are only six orders of magnitude from the radius of our galaxy to the radius of the observable universe (\SI{\approx e27}{\metre}) and infinitely many beyond that, he reaches a conclusion that it is unlikely for two civilizations to arise within the same observable universe. Brian Lacki has suggested an improvement to Tegmark's model, in which the log-uniform prior is replaced with a bounded log-log-uniform prior \cite{lacki2016loglog}.

Our argument shares the same broad outline. But rather than starting with a very abstract prior representing initial radical uncertainty over more than $10^{100}$ orders of magnitude, we used two different methods to provide a prior that captured the existing scientific uncertainties of tens or hundreds of orders of magnitude. We have seen how this is more than enough make an empty observable universe plausible {\em ex ante} (dissolving the Fermi paradox), and quite likely once we account for the Fermi observation.

\section{Updating the factors}
So far we have looked at how the Fermi observation affects our credence in $N$. We can go further than this and examine how it affects our credence in each of the Drake parameters. Updating on the Fermi observation reduces the expectation of all the parameters. However, parameters with broad distributions (those with the most uncertainty) tend to have their expectation reduced far more than parameters with tight distributions (see Supplement IV).

All the observations we consider have a strong effect on our estimates for $f_l$, a substantially weaker effect on our estimates for $L$, and almost no effect on our estimates of the more certain astrophysical factors. As we can see in Table~\ref{tab:conditioning}, the observations reduce the median for $f_l$ by between a factor of 7 and factor of $10^{37}$, while the median for $L$ is only reduced by a factor between 1 and 2. 

Given the state of scientific uncertainty about the Drake parameters and the Fermi observation, the default guess should hence be that the low-probability term is likely in the past ($f_l$) rather than the future ($f_c, L$). The Fermi observation thus provides only very weak evidence about whether we will soon go extinct or whether interstellar communication or travel is impossible. Instead, the observation mainly just increases our credence that life is rare.

This conclusion is quite robust to changing the log-uncertainties of the factors (it remains as long as most uncertainty is in the past factors) or their distribution shape (using log-normals instead of log-uniform distribution has no effect). The conclusion can be changed if we reduce the uncertainty of past terms to less than just 7 orders of magnitude, or if the $f_c$ factor turns out to be {\em radically} uncertain.

\section{Conclusion}
We have seen that a Fermi paradox arises if we combine a high and extremely confident prior for the number of civilizations in our galaxy with the absence of evidence for their existence. The high confidence that causes this clash typically results from applying a Drake-like model using point estimates for the parameters. These estimates, however, make implicit knowledge claims about processes (especially those connected with the origin of life) which are untenable given the current state of scientific knowledge.

When we take account of realistic uncertainty, replacing point estimates by probability distributions that reflect current scientific understanding, we find no reason to be highly confident that the galaxy (or observable universe) contains other civilizations, and thus no longer find our observations in conflict with our prior probabilities. We found qualitatively similar results through two different methods: using the authors' assessments of current scientific knowledge bearing on key parameters, and using the divergent estimates of these parameters in the astrobiology literature as a proxy for current scientific uncertainty.

When we update this prior in light of the Fermi observation, we find a substantial probability that we are alone in our galaxy, and perhaps even in our observable universe (53\%--99.6\% and 39\%--85\% respectively).  'Where are they?' --- probably extremely far away, and quite possibly beyond the cosmological horizon and forever unreachable.

\enlargethispage{20pt}



\paragraph{Authors’ Contributions:} AS, TO. and ED wrote the paper together; AS performed the literature review and simulations.

\paragraph{Competing Interests:} The authors declare no conflict of interest.

\paragraph{Funding:} This project has received funding from the European Research Council (ERC) under the European Union's Horizon 2020 research and innovation programme (grant agreement No 669751). This paper reflects only the view of the authors: the ERCEA is not responsible for any use that may be made of the information it contains.

\paragraph{Acknowledgments:} We would like to acknowledge Andrew Snyder-Beattie, Stuart Armstrong, Carl Shulman and many others at FHI for stimulating and useful conversations. We also had great input from Olle H\"aggstr\"om, Vilhelm Verendel, Karim Jebari, Pontus Strimling, Allan Penny, Stephen Webb, David Brin, Geoff Marcy, and Milan \'Cirkovi\'c.



\bibliography{fermibib}

\end{document}